\documentclass[12pt]{article}
\usepackage{graphicx}
\usepackage{a4}
\usepackage{amssymb}
\usepackage{graphicx}
\usepackage{epsf}
\usepackage{cite}
\newcommand{\be}{\begin{equation}}
\newcommand{\ee}{\end{equation}}
\newcommand{\bea}{\begin{eqnarray}}
\newcommand{\eea}{\end{eqnarray}}

\def\myp{Myers-Perry}
\begin{document}

\def\C{{\mathbb{C}}}
\def\R{{\mathbb{R}}}
\def\s{{\mathbb{S}}}
\def\T{{\mathbb{T}}}
\def\Z{{\mathbb{Z}}}
\def\W{{\mathbb{W}}}
\def\Bbb{\mathbb}
\def\BZ{\Bbb Z} \def\BR{\Bbb R}
\def\BW{\Bbb W} 
\def\BM{\Bbb M} 
\def\e{\mbox{e}}
\def\BC{\Bbb C} \def\BP{\Bbb P}
\def\CP{\BC\BP}

\begin{titlepage}
\title{Brane World Black Rings.}
\author{}
\date{Anurag Sahay\thanks{\noindent ashaya@iitk.ac.in}, 
Gautam Sengupta, \thanks{\noindent sengupta@iitk.ac.in}, \\
\vspace{1.0cm}
Department of Physics, \\ Indian Institute of Technology,\\
Kanpur 208016, India.}
\maketitle
\abstract{ Five dimensional neutral rotating black rings are described from a Randall-Sundrum
brane world perspective in the bulk black string framework. To this end we consider
a rotating black string extension of a five dimensional black ring into the bulk 
of a six dimensional Randall-Sundrum brane world with a single four brane. The bulk
solution intercepts the four brane in a five dimensional black ring with the usual
curvature singularity on the brane. The bulk geodesics restricted to the plane of rotation of 
the black ring  are constructed and their projections on the four brane match 
with the usual black ring geodesics restricted to the same plane.
The asymptotic nature of the bulk geodesics are elucidated with reference to a bulk 
singularity at the AdS horizon. We further discuss the description of a brane world black ring 
as a limit of a boosted bulk black 2 brane with periodic identification. }

\vskip .2in
April 2007\\
\end{titlepage}

\section {Introduction.}  \label{one}
\noindent

Higher dimensional spacetimes are now an essential aspect of effective field theories arising from fundamental theories of quantum gravity. The general assumption implicit in such constructions was that the extra spatial dimensions are compactified to ultrashort length scales. Hence quantum gravity effects were relegated to very high energy scales. However in recent years the exciting possibility of {\it low scale quantum gravity effects} in the {\it brane world models} have inspired
considerable interest and interesting phenomenological consequences \cite {add,lyk,pheno}. The {\it brane world scenario} envisaged the gauge sector of the fundamental interactions to be restricted on a smooth
codimension one hypersurface ( refered to as a {\it brane}) embedded in a higher dimensional space-time and the electroweak scale as the fundamental scale. The usual four dimensional Planck scale was then a derived scale.
In particular the Randall-Sundrum models and their variants based
on a warped non factorable compactification geometry in a bulk Anti deSitter ( AdS) space time offered a partial resolution to the vexing {\it hierarchy problem} \cite {rs12}. Although the analysis was valid in a linearized
framework a full non linear study from a supergravity perspective confirmed the conclusions
and their extension to any Ricci flat geometry on the brane\cite {dick, gibs}. 

For consistency the brane world scenario requires generic
four dimensional gravitational configurations on the brane to arise from a higher dimensional bulk.  The 
investigation of black hole configurations in this context has been an exciting aspect of the study of brane world gravity \cite {dick}.  Such a black hole on the brane is expected to be a configuration extended in the bulk. Chamblin, Hawking and Reall  \cite {chr} attempted the description of a Schwarzschild black hole in a typical single three brane five dimensional Randall-Sundrum brane world as a bulk black string. This reproduced the usual Schwarzschild singularity on the brane but additionaly was also singular at the AdS horizon far away from the three brane. Although a pathology, this singularity was possibly a linearization artifact and could be shown to be a mild p-p curvature singularity. The bulk black string
was subject to the usual instabilities against long wavelength perturbations \cite {greg,greg1} and was expected to pinch off to a cigar geometry before reaching the AdS horizon. However the issue of stability is contentious and for sphericaly symmetric solutions it was shown that a more likely scenario is a  transition to a non uniform black string \cite {maeda}

In an earlier article \cite {gs}we have generalized the construction of Chamblin et. al.  \cite {chr}to consider rotating black holes in a five dimensional single three brane RS brane world. The bulk configuration proposed was  a five dimensional rotating black string which intercepted the three-brane in a four dimensional rotating black hole described by a Kerr metric on the three brane.  It was found that the Kerr solution too was singular at the AdS horizon apart from the usual ring singularity on the brane.  The asymptotics of the equatorial geodesics at the AdS horizon also indicated a p-p curvature singularity although an explicit
determination was computationaly intractable. There have been other approaches to brane world black holes including numerical studies for off brane metrics and a Hamiltonian constraint approach to charged black holes \cite {gian,exrefs,kim,cbh,num,hd, morebrane}. In lower dimensions exact studies of brane world black holes \cite {emp} involving the AdS C-metric have indicated that the bulk solutions are regular everywhere emphasizing that the bulk singularity in higher dimension is possibly a linearization artifact. However absence of
exact bulk metrics in higher dimensions requires a linearized  approach and the black string framework is hence physicaly relevant in this context in spite of such  a bulk singularity.  

The brane world constructions must be embedded in an appropriate string theory for consistency,
requiring the generalizations of these models to higher dimensions. The generalization of the Randall-Sundrum construction and its variants to higher dimensions with a single space like AdS direction and an appropriate codimension one brane is straightforward.  Additionaly this may easily be extended to include the full non linear extensions of a Ricci flat metric \cite {chr1,gid,pps}.  In higher dimensions
also the consistency of such brane world constructions require that gravitational
configurations arise from appropriate bulk scenarios. In particular this applies to higher dimensional black holes on the codimension one brane. In this context in an earlier article \cite {gs1}we had  described the
N dimensional rotating Myers-Perry \cite {myers}black hole on a single (N-1) brane in a (N + 1) dimensional RS brane world. The bulk solution in this case was a  (N+1) dimensional rotating black string extended in the AdS direction transverse to the (N-1) brane. Analysis of equatorial geodesics again indicated a p-p curvature singularity in the bulk apart from the usual extended singularity on the (N-1) brane.

In the recent past there has been remarkable and surprising progress in understanding higher dimensional black holes. In particular it has been realized that the {\it no hair} and the {\it uniqueness} theorems  are much less restrictive in higher dimensions \cite {cai}. In four dimensions the no hair theorem characterizes any stationary asymptoticaly flat black hole solution of Einstein-Maxwell system only by their mass, angular momentum and conserved charges whereas the uniqueness theorem forbids event horizons of non spherical toplogies. However the discovery \cite {reall} in five dimensions of an asymptoticaly flat stationary black hole
solution with a non spherical ring like $S^2\times S^1$ horizon topology with the possibility of  dipole charges, showed that higher dimensional black holes posess remarkably distinctive properties. 
The static black ring solution \cite {empring} was first obtained through the Wick rotation of a neutral solution of
an Einstein-Maxwell system \cite {empchamb} although they involved conical singularities. However the stationary solution rotating in the $S^1$ direction was regular everywhere except the usual curvature singularity. For fixed mass the angular momentum of the black ring was bounded below and for a certain range of parameters two black rings and a usual five dimensional
rotating Myers-Perry black hole all with the same mass and spin coexist.  The charged versions of these black rings were first obtained in the framework  of D=5 heterotic supergravity \cite {elvang} and fully supersymmeric three charged black ring solutions in D=5 followed later from compactifications of black supertubes in D=10 \cite {susyring, revring}. It was seen that these black rings could also support gauge dipoles independent of the conserved gauge charges entailing an infinite non uniqueness and violating the {\it no hair theorem}\cite {dipring}. 

As emphasized earlier, for consistency of the brane world scenario it is imperative that
gravitational configurations like black holes on the brane should arise from appropriate bulk solutions.  In this context it is but natural to investigate possible bulk configurations in a higher
dimensional brane world scenario which would describe five dimensional black rings on the
brane. This is especialy relevant for the neutral rotating black rings as they are Ricci flat and hence satisfy the criteria for embedding in higher dimensional Randall-Sundrum brane worlds. Naturaly the absence of exact solutions in higher dimensions require the usual linearized framework to analyse this question.
The black string approach is especialy relevant in this context to highlight the physical aspects of such an embedding although it suffers from singular pathologies which are possibly linearization artifacts. 

In this article we address this issue and show that it is possible to consistently embed the five dimensional black ring solution on a single four brane in a (5 + 1) dimensional Randall-Sundrum brane world.  Following the black string approach we consider a six dimensional bulk rotating black string extension of the five dimensional black ring. This bulk configuration intercepts the four brane in a five dimensional rotating black ring. In what follows after a brief review of neutral rotating black rings, we obtain their geodesic equations in the plane of the ring analogous to the equatorial plane of black holes with spherical topologies. We further investigate the asymptotic behaviour of both the null and the timelike geodesics in this plane to elucidate the restricted causal structure of the black ring space time. In section three we consider
a bulk rotating black string extension of a five
dimensional neutral rotating black ring in a six dimensional  RS brane world with a single four brane. The bulk black string intercepts the four brane in a five dimensional black ring with the usual spacelike curvature singularity on the brane. Additionaly
a curvature singularity also appears at the AdS horizon far away from the four brane.
Following the
description of a black ring as a boosted black string with periodic identification in a certain limit,
the bulk solution may be described as a boosted black two brane with the same periodic identification. We then construct the six dimensional bulk geodesics 
in the plane of rotation of the ring and show that their projections on the four brane reproduces
the usual five dimensional black ring geodesics in the same plane. To study of the nature of the
pathological singularity at the AdS horizon we further investigate the late time asymptotics
of these geodesics. It is shown that the curvature remains finite along unbound geodesics which
reach the AdS horizon. We also discuss the possibility of the bulk solution to pinch off before reaching the AdS horizon due to the usual instabilities and comment on the possible stable solution in the light of the analysis outlined in \cite {greg1} and \cite {maeda}. In the last section we provide a  summary of our analysis and results and also discuss certain future open issues in this area.   

\section {The Rotating Neutral Black Ring .} \label{two}
In this section we first briefly review the neutral rotating black ring and elucidate the nature of the adapted coordinate system .  We then construct the black ring geodesics restricted to the plane of rotation of the ring which is analogous to the equatorial geodesics in solutions with a spherical topology. Furthermore we analyse the geodesic equations to study the nature of the radial orbits for this plane and their asymptotics. The static neutral black ring was originally discovered through a Wick rotation of certain Kaluza Klein C metrics decribing neutral bubbles\cite {empchamb}. These involved conical singularities and consequent deficit angles leading to either cosmic string defects joining these singularities or deficit membranes. However an analytic continuation  led to the original neutral rotating black ring solution which was a five dimensional asymptoticaly flat black hole with a ringlike $S^2\times S^1$  horizon topology, regular everywhere except at a spacelik
 e curvature singularity. The original solution was further refined through appropriate factorizable choice of certain functions appearing in the metric \cite {elvang, susyring,revring,teo}.  The rotating black ring in equlibrium was parametrized by a dimensionless reduced angular momentum $j=\frac {27\pi}{32 G}\frac {J^2}{M^3}$ which was bounded from below for a fixed mass. It could be shown that in the range $\frac {27}{32}\leq j^2 < 1$ there existed one Myers-Perry black hole with spherical topology and two black rings with identical mass and angular momenta, in direct violation of the black hole uniqueness theorem.

\subsection{Black Ring Metric}

The metric of the neutral rotating five dimensional black ring in a specific adpated coordinate
system which is obtained from the foliation of space-time in terms of the equipotentials of certain
1-form and 2-form gauge potentials is ,\cite{revring}

\begin{eqnarray}
ds^2&=&-\frac{F(y)}{F(x)}\left(dt-C\: R\:\frac{1+y}{F(y)}\:
d\psi\right)^2 \nonumber \\ [2mm]
&&+\frac{R^2}{(x-y)^2}\:F(x)\left[
-\frac{G(y)}{F(y)}d\psi^2-\frac{dy^2}{G(y)}
+\frac{dx^2}{G(x)}+\frac{G(x)}{F(x)}d\phi^2\right]\,,
\end{eqnarray}
where the functions
\begin{equation}
F(\xi)=1+\lambda\xi,\qquad G(\xi)=(1-\xi^2)(1+\nu\xi)\,,
\end{equation}
and
\begin{equation}
C=\sqrt{\lambda(\lambda-
\nu)\left( \frac{1+\lambda}{1-\lambda}\right) }\,.
\end{equation}

Here $R$ is a length scale which may be interpreted as the radius of the ring in some limit \cite {revring}
and the two dimensionless parameters
$\lambda$ and $\nu$ which are related to the shape and the rotation velocity of the ring lie in the range 
\begin{equation}
0< \nu\leq\lambda<1\
\end{equation}.
The range of the spatial co-ordinates $(x,y)$ are required to be,
\begin{equation}
-1\leq x\leq +1\,,\qquad-\infty\leq y\leq -1\,.
\end{equation}
respectively. 

The constant $y$ hypersurfaces are nested deformed solid toroids with topology $S^2 \times S^1$, whereas the coordinate $x$ is like a direction cosine, $x=+1$ points to the interior of the ring
and $x=-1$ points to the region outside the ring. The solution is a stationary axisymmetric solution with rotation in the $\psi$ direction, and admits $t$, $\phi$, and $\psi$ Killing isometries. 

In order to avoid conical singularities at the fixed points $x=-1$ and $y=-1$ of the Killing isometries $\partial_{\phi}$ and $\partial_{\psi}$ the co-ordinates $\psi$ and $\phi$ require to be identified with the equal periods
\begin{equation}
\Delta\psi=\Delta\phi=4\pi\frac{\sqrt{F(-1)}}{|G'(-1)|}=
2\pi\frac{\sqrt{1-\lambda}}{1-\nu}\,.
\end{equation}
Furthermore the requirement that the orbits of the isometry $\partial_{\phi}$ shows no deficit angles at $x=+1$ lead to the condition
\begin{equation}
\lambda=\frac{2\nu}{1+\nu^2}
\end{equation}
The co-ordinates $(x,\phi)$ parametrize a two-sphere $S^2$, the co-ordinate $\psi$ parametrizes a circle $S^1$ and the solution describes a black ring having a regular horizon of topology $S^1\times S^2 $ and rotating in the $S^1$  plane. However the horizon geometry is not a simple product of $S^2$ and $S^1$ as the two sphere $S^2$ is deformed there and the deformation grows away from the horizon.

The metric reduces to a conventional five dimensional \myp\	black hole with rotation in a single plane if, instead of (7), we consider the limit, $R\to0$,  $(\lambda,\nu) \to1$ and the parameters 
\begin{equation}
m=\frac{2R^2}{1-\nu}\,,\qquad a^2=2R^2\frac{\lambda-\nu}{(1-\nu)^2}\,,
\end{equation}
are held constant. In this case the co-ordinates $(x,\phi,\psi)$ characterises a three-sphere $S_3$ which is a regular horizon of a five dimensional  Myers-Perry black hole. The ergosphere and the event horizon of the black ring are located at $y=-1/\lambda$ and $y=-1/\nu$ respectively. At $y=-\infty$ there is a spacelike curvature singularity inside the horizon. Asymptotic infinity is reached as $(x,y)\to -1$.
 
The ADM mass and angular momentum are given as
\begin{eqnarray}
M&=&\frac{3\pi R^2}{4G }\frac{\lambda}{1-
\nu}\\
J&=&\frac{\pi R^3}{2G }\frac{\sqrt{\lambda(\lambda-
\nu)(1+\lambda)}}{(1-\nu)^2}.
\end{eqnarray}
 
The curvature squared for the black ring spacetime is computed to be,
\begin{equation}
R_{\mu\nu\rho\sigma}R^{\mu\nu\rho\sigma}=\frac{6{\nu}^2(1+{\nu}^2)^2Q(x,y)}{R^4(1+{\nu}^2+2\nu x)^6}(x-y)^4,
\end{equation}
where $Q(x,y)$ is a poynomial of degree six in $x$ and $y$.  Hence there is  a spacelike curvature singularity at $y=-\infty$ inside the event horizon.
In terms of the \myp\	 co-ordinates $(t,r,\theta,\psi,\phi)$ the difference ($x-y$) goes like $1/r^2$ at large $r$, {\it i.e.} towards spatial infinity, so that the curvature squared goes as
\begin{equation}
R_{\mu\nu\rho\sigma}R^{\mu\nu\rho\sigma} \sim \frac{1}{r^8}
\end{equation}
as obtained in the case of five dimensional \myp\	black hole. 

The rotating black ring in the limit of large radius $R$ may be described after appropriate coordinate redfinitions as a Schwarzschild black string boosted and periodically identified along the translation invariant  direction with a period $2\pi R$ \cite {elvang,revring,dipring}. The black string metric is given as
\begin{equation}
ds^2=dw^2-(1-\frac{r_0}{r})dt^2+(1-\frac{r_0}{r})^{-1}dr^2+r^2d{\Omega}_2^2,
\end{equation} where the horizon is at $r=r_0$ and $w$ is the translation invariant direction.
The parameter $\nu= {r_0}/{R}$ is seen to correspond to the thickness of the ring or the ratio of the radius of the $S^2$ at the horizon and the ring radius $R$ .
The ratio $ {\lambda}/{\nu}$ then measures the speed of rotation of the ring in the $S^1$ direction and the coordinate $\psi={w}/{R}$  corresponds to a redefined translation invariant direction of the black string which is periodically identified as $w=w +2\pi R$. The speed of rotation is related to the local boost velocity given by $\sqrt {1-({\nu}/{\lambda})}$ and reduces to $\sqrt {1-({\nu^2}/{2})}$ for the black ring space time to exclude any conical singularities. \cite {elvang}

\subsection{Black Ring Geodesics}

The first order geodesic equations may be derived using the canonical framework \cite {wald} from the Lagrangian
\begin{equation}\label{lagr}
{\cal L} = \frac{1}{2}g_{\mu\nu}\dot x^{\mu}\dot x^{\nu},
\end{equation}
here $\mu, \nu=0...4$ and
the covariant components of the metric tensor are as defined in the previous section and $\dot x^{\mu}=dx^{\mu}/d\rho$ with the affine parameter $\rho={\tau}/{m}$ \cite {frolov}for time like geodesics, $\tau$ being the proper time and $m$ the mass of the particle. Consequently, for both time like and null geodesics the momenta are  $p^{\mu}=\dot x^{\mu}$. The covariant momenta may be directly  obtained from the Lagrangian and are given as $p_{\mu}=g_{\mu\nu}\dot x^{\mu}$.  The norm of the conjugate momenta is then given as,
\begin{equation}
 g^{\mu\nu}p_{\mu}p_{\nu}=-{\epsilon}m^2
\end{equation}
where $g^{\mu\nu}$ are the contravariant components of the black ring metric and $\epsilon=(0, 1)$ for null and time like geodesics respectively.

The black ring spacetime admits three Killing isometries generated by the vector fields $\partial_{t}
$, $\partial_{\psi}$, and $\partial_{\phi}$ corresponding to time translation and the two rotation isometries in the coordinates $\phi$, $\psi$. These isometries provide three conserved conjugate momenta, $p_t=-E$, $p_{\psi}=\Psi$, $p_{\phi}=\Phi$.
We consider the geodesics restricted to the plane of rotation of the black ring, outside the ring, \textit{i.e}, $x=-1$. It is analogous to an equatorial plane in the spherical case in the sense that it is reflection symmetric and hence geodesics in it with zero initial velocity in the transverse $x$ direction will continue to remain in the plane. The plane $x=-1$ being a fixed point of the $\partial_{\phi}$ isometry, the $g_{\phi\phi}$ component of the metric tensor goes to zero smoothly there. 
The geodesic equations of motion in the equatorial plane for the $t$ and $\phi$ directions are obtained directly from the conserved conjugate momenta. These turn out to be as follows:
\begin{eqnarray}
\frac{dt}{d\rho}&=&\frac{1-\lambda}{1+{\lambda}y}\left(\frac{C^2}{(1-\lambda)^2}\frac{(1+y)^4}{(1+{\lambda}y)G(y)}+1\right)E-\frac{C(1+y)^3}{R(1-\lambda)(G(y)}\Psi\\
\frac{d\psi}{d\rho}&=& \frac{CR(1+y)^3}{(1-\lambda)G(y)}E-\frac{(1+y)^2(1+{\lambda}y)}{R^2(1-\lambda)G(y)}\Psi
\end{eqnarray}
The form of the $y$ equation for geodesic motion in the equatorial plane is obtained directly from 
eqn. (15) to be,
\begin{equation}
(\frac{dy}{d\rho})^2+g^{yy}\left( g^{tt}E^2-2g^{t\psi}E\Psi+g^{\psi\psi}{\Psi}^2+{\epsilon}m^2\right)=0 ,
\end{equation}
where
$g^{yy}=1/g_{yy},
g^{tt}=g_{\psi\psi}/D, 
g^{\psi\psi}=g_{tt}/D,
g^{t\psi}=-g_{t\psi}/D $ and $D=g_{tt}g_{\psi\psi}-{g_{t\psi}}^2$.\\

Thus, the $y$ equation may be expressed as
\begin{eqnarray}
{\dot{y}}^2=&-& \frac{(1+y)^3}{(1-\lambda)^2R^2}\left (\frac{C^2(1+y)^3+(1-\lambda)^2(1+\nu y)(1-y)}{F(y)} E^2  \right.\nonumber \\ 
&-& \left. \frac{2C(1+y)^2}{R}E{\Psi}+\frac{(1+\lambda y)(1+y)}{R^2}\Psi^2\nonumber \right.\\
&-& \left. \epsilon(1-\lambda)(1+\nu y)(1-y)m^2  \right )  
\end{eqnarray}
where $\epsilon=(0,1)$ for null and timelike geodesics respectively. It should be noted that the co-efficient of $E^2$ in the r.h.s of the above equation remains finite and smooth at the ergosphere, $y=-1/\lambda$, even though the function $F(y)$ in the denominator vanishes. The eqn (19) should be compared with that appearing in \cite {vir} for the null geodesics in the plane of the ring, where a 
a certain normalization of the metric components have been chosen at asymptotic infinity.

The $y$ co-ordinate ranges over the plane of rotation of the ring from the curvature singularity to asymptotic infinity and the above equation is analogous to particle motion in a central potential
\begin{equation}
{\dot{y}}^2+V_{eff}(y;E,\Psi)=0
\end{equation}

Towards asymptotic infinity, $(x,y) \to -1$, the effective potential for time like geodesics tends to
\begin{equation}
V_{eff}(y;E,\Psi)\to -\frac{2(1-\nu)}{R^2(1+\lambda)}{\eta}^3(E^2-m^2),
\end{equation}
where $\eta$ tends to $0$ towards asymptotic infinity and is given by $\eta=-(1+y)$.

Unbound time like geodesics can exist only when $E^2-m^2>0$  in which case the effective
potential $V_{eff}$ is negative at large distances and approaches zero at asymptotic
infinity $(x,y=-1)$. For the case $E^2<m^2$ only bound geodesics exist, in the sense
that such geodesics do not reach upto asymptotic infinity.  
Stable bound orbits are bound orbits which do not end up in the singularity.
It is common knowledge that stable bound orbits
do not occur in a higher dimensional central potential, even in the case of Newtonian gravity.  Thus it is expected that such orbits must be excluded from higher dimensional black hole space times. This was
explicitly shown for the equatorial geodesics of a five dimensional Myers-Perry black hole in 
\cite {frolov}. 
This conclusion is expected to also hold for the class of geodesics restricted to the plane of rotation of the ring being considered here.  Their existence is indicated by the 
presence of stable circular orbits. For circular
orbits, we have the condition
\begin{equation}
V_{eff}(y=y_c)=0\ ,\qquad \frac{\partial{V_{eff}(y)}}{\partial{y}}{\mid}_{y=y_c}=0
\end{equation}
where $y=y_c$ is the `radius' of the circular orbit.The condition for stability of the
circular orbit is
\begin{equation}
\frac{{\partial}^2 V_{eff}}{\partial y^2}{\mid}_{y=y_c}>0 .
\end{equation}
\begin{figure}[htb]
\centering
\includegraphics[height=3.5in]{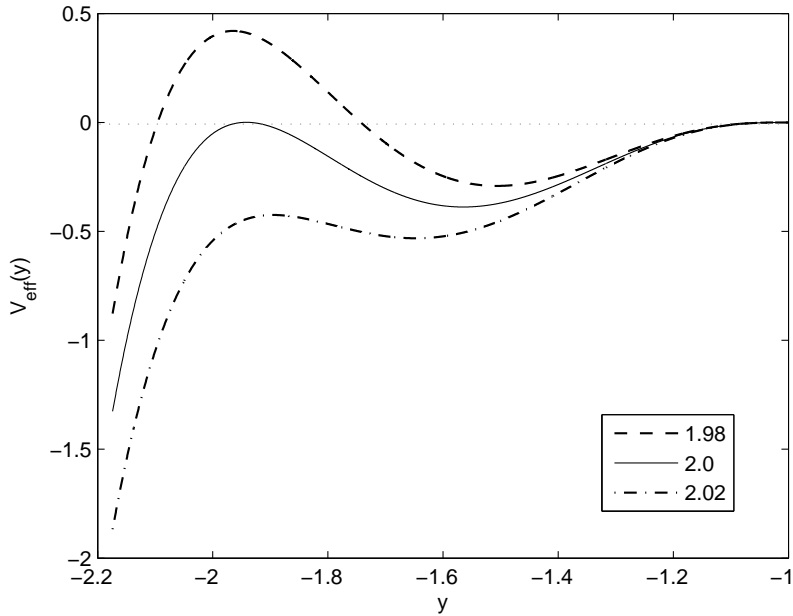}
\label{Fig. 1}
\caption[short] { Plot of black ring effective potential for $\nu=0.46,$  $L=4.40145$ and three different values of $E$ as indicated in the box. Motion is allowed only in the region where
$V_{eff} < 0$. The constants $m=R=1$. It is apparent that there are no stable bound orbits. $E=2.0$ is close to having an unstable circular orbit, whereas for $E=2.02$ there are no inaccessible regions. Since $E > 1$ all the three curves exhibit unbounded orbits. The case for $E < 1$ shows an exactly similar behaviour as regards the bound orbits.} 
\end{figure}
\vskip 5pt

We get two simulataneous biquadratic equations in $E$ and $\Psi$ from Eq(22) which
can be solved in terms of the radius $y_c$ for a black ring of specific $\nu$. These
values of $E_c$ and ${\Psi}_c$ can be then substituted into (23) to obtain a function
of $y_c$ for a specific black ring \cite {frolov}.
It is difficult to interpret the analytic expressions for $E_c,{\Psi}_c$ and that
of Eq.(23) in terms of $y_c$.  However, numerical plots
have been obtained in Fig. 1 for the effective potential $V_{eff}(y)$ against $y$
which clearly shows that stable bound orbits are ruled out both for $E^2 >m^2$ and $E^2 < m^2$ .

\section { Brane World Black Ring}\label {three}
In this section we very briefly outline the construction of the Randall-Sundrum braneworld with
a single (N-1)-brane in (N+1) dimensions with a single AdS direction transverse to the
brane. We then consider the specific case of the five dimensional neutral rotating black ring on a four brane in a (5+1) dimensional Randall-Sundrum braneworld with a single AdS direction transverse to the brane hypersurface. We propose that the appropriate bulk description is provided by a six dimensional rotating black string extension of the five dimensional rotating black ring. The intercept
of the bulk solution on the four brane is a five dimensional black ring with the usual curvature singularity on the brane hypersurface although an additional bulk singularity also appears at the AdS
horizon. We also compute the six dimensional bulk geodesics restricted to the plane of rotation of the black ring.  The projection of these bulk geodesics on the four brane reduces to the appropriate class of black ring geodesics on the four brane hypersurface. The $y$ orbits for the bulk solution which reach the AdS horizon are then  analyzed using the geodesic equation to elucidate the natuer of the
bulk singularity at the AdS horizon. It is seen that the curvature remains finite  at the AdS horizon along the unbounded indicating the presence of a mild p-p curvature singularity.

\subsection {Black Ring in a RS Brane World}

The bulk metric for single brane RS brane world in (N +1) dimensions,  with one transverse AdS direction to the (N-1) brane is as follows; \cite {emp,gid} 
\be
ds^2=g_{mn}dx^mdx^n={l^2\over z^2}\big [ g_{\mu\nu}dx^{\mu}dx^{\nu} + dz^2
\big ].
\ee
Here  $\mu,\nu = 0\ldots (N-1) $ and $m,n=0\ldots (N) $ and
$l$ is the AdS length scale.  The transverse coordinate $z=0, \infty$ are the conformal infinity
and the AdS horizon respectively. The actual RS braneworld geometry is obtained by removing the small $z$ region at
$z=z_0$ and glueing a mirror copy of the large $z$ geometry at the location of the (N-1) brane which ensures $Z_2$ reflection symmetry. The
resulting topology for the double brane RS scenario is 
essentialy $R^N \times {{S^1}\over {Z_2}}$ and in the single brane variant considered here the $S^1$ direction
is essentialy decompactified with the second regulator brane being at $z=\infty$.
The discontinuity  of the extrinsic curvature at
the $z=z_0$ surface corresponds to a thin distributional source of stress-energy. From the Israel junctions conditions this may be interpreted as
a relativistic  (N-1) brane (smooth domain wall) with a corresponding tension \cite{gid, emp}.
The orginal RS model sliced the AdS space-time
both at $z=0$ and $z=l$ and inserted two (N-1) branes with $Z_2$ reflection
symmetry at both hypersurfaces. The Israel junction conditions  then required a
negative tension for the brane at $z=l$. The variant considered here may be obtained
from the original RS model by allowing the negative tension brane to approach the
AdS horizon at $z=\infty$ . Although we focus here only on the single brane RS
model for convenience, our construction
may be generalized to the original RS model with double branes in a straightforward manner.

The Einstein equations in (N+1) dimensions with a negative cosmological
constant continue to be satisfied for any metric $g_{\mu\nu}$ which is
Ricci flat. The curvature of the modified metric now satisfies
\be
R_{pqrs}R^{pqrs}= {{2N(N+1)}\over l^4} +{z^4\over
l^4}R_{\mu\nu\lambda\kappa}R^{\mu\nu\lambda\kappa}
\ee where $(p,q)$ runs over (N +1) dimensions and $(\mu , \nu)$ over the N dimensions of
the brane world volume. The perturbations of the (N+1) dimensional metric around a Ricci flat
background are now normalizable modes peaked at the location of the (N-1) brane.

Having provided this brief introduction to the single brane RS model in (N+1) dimensions
we now specialize to N=5 and consider the bulk description of a five dimensional neutral 
rotating black ring on  the four brane in a six dimensional RS braneworld. To this end we consider a bulk six dimensional black string extension of the five dimensional rotating neutral black ring in the bulk. The black ring being a Ricci flat space-time the bulk black string extension automaticaly satisfies the Einstein equation \cite {gid}
For a reflection symmetric four brane hypersurface fixed at $z=z_0$ we may introduce the co-ordinate $w=z-z_0$. The bulk metric on either side of the domain wall may now be expressed
as
 
\begin{eqnarray}
ds^2&=& \frac{l^2}{(z_0+\vert w \vert)^2}\left[dw^2 -\frac{F(y)}{F(x)}\left(dt-C R\frac{1+y}{F(y)}\:
d\psi\right)^2 \right . \nonumber \\ [2mm]
&+&\left .\frac{R^2}{(x-y)^2}F(x)\left(
-\frac{G(y)}{F(y)}d\psi^2-\frac{dy^2}{G(y)}
+\frac{dx^2}{G(x)}+\frac{G(x)}{F(x)}d\phi^2\right)\right]\	
\end{eqnarray} 
where $-\infty <w<\infty$ and the domain wall is located at $w=0$.

The induced metric on the four brane at $z=z_0$ may be recast into the black ring form by suitably rescaling the coordinates and the parameters. The ADM mass and angular momentum as measured on the brane,  scaled by the conformal warp factor, are then given as
\begin{eqnarray}
M_*=\left( \frac{l}{z_0}\right)^2 M \,,\qquad J_*=\left( \frac{l}{z_0}\right) ^3J.
\end{eqnarray}
where $M, J$ are the bulk parameters.

\noindent The curvature squared for the bulk black string is computed to be;
\begin{equation}
R_{jklm}R^{jklm}= \frac{1}{l^4}\left[ 60 +\frac{6(1+{\nu}^2)^2{\nu}^2Q(x,y)}{R^4(1+{\nu}^2+2\nu x)^6}z^4(x-y)^4\right] 
\end{equation}
Following Eq(12), towards spatial infinity on the brane the curvature squared behaves as
 \begin{equation}
R_{jklm}R^{jklm}\sim \frac{z^4}{r^8}.
\end{equation}
The curvature invariant diverges at the spacelike singularity on the brane at $y=-\infty$. Additionaly, it is also seen to diverge at the AdS horizon $z=\infty$ for finite $r$. As mentioned earlier, such a singularity seems to be a artifact of the linearized approximation. In order to further investigate this issue we need to study the geodesics and their behaviour at the AdS horizon. 

As mentioned earlier the neutral rotating black ring maybe described in a certain  limit as a Schwarzschild black string boosted in the translationaly invariant direction and identified periodicaly. In the braneworld construction that we have developed, this reduces to a six dimensional bulk black two brane boosted along the extended direction on the four brane and identified periodically. 
In the 5+1 dimensional brane world Eq. (13) generalizes to;
\begin{equation}
ds^2=\frac{l^2}{z^2}\left[  dz^2+dw^2-(1-\frac{r_0}{r})dt^2+(1-\frac{r_0}{r})^{-1}dr^2+r^2d{\Omega}_2^2 \right] 
\end{equation}
Here $u$ is the translation invariant direction of the black string along the brane hypersurface and $z$ describes the transverse direction. Apart from the conformal factor the coordinate $z$ is a spectator dimension
and  hence we have a six dimensional bulk Schwarzschild black two brane boosted along a translation invariant direction $w$ and periodicaly identified as  $w\sim w+2\pi R$. This bulk black two brane in the limit of large boost velocity and a large periodicity $R$ intercepts the four brane in a fast spinning thin five dimensional neutral black ring of large radius $R$ with the usual curvature singularity on the brane. This is obvious as the boost does not involve the transverse $z$ direction and the limit of large radius and high boost velocity are $z$ independent.  So in this limit after periodic identification the event horizon has $S^2\times S^1\times R$ topology extended in the bulk and periodic in the coordinate $w$ on the four brane.

\subsection{The Brane World Geodesics.}
The geodesic equations for the the bulk spacetime may be obtained as earlier from the Lagrangian
\begin{equation}
{\cal L}=\frac{1}{2}={g}_{jk}\dot x^{j}\dot x^{k}
\end{equation}
where $g_{jk}$ are the covariant components of the 5+1 dimensional metric as in eqn. (24) and $j,k=0\ldots5$. Also $\dot x={dx}/{d\rho}$ and on time like geodesics the affine parameter $\rho= {\tau}/{m}$. Accordingly we have
$p^j=\dot x^j,$
$p_j=g_{jk}\dot x^k$ and
\begin{equation}
{g}^{jk}p_{j}p_{k}=-{\epsilon}m^2\\
\end{equation}
where $\epsilon=0,1$ for null and time like geodesics respectively.

The $z$ equation for geodesic motion is obtained from the Lagrangian as 
\begin{equation}
\frac{d}{d\rho}\left( \frac{1}{z^2}\frac{dz}{d\rho}\right) =\frac{{\epsilon}m^2}{zl^2}.
\end{equation}
The solution for null geodesics is either $z=$constant or 
 \begin{equation}
z=-\frac{z_1 l}{m\rho}.
\end{equation}
For timelike geodesics the solution is
\begin{equation}
z=-z_1cosec({\rho}m/l).
\end{equation} 
Here $m$ is the particle mass for timelike geodesics and we should set $z_1/m$=constant for the null geodesics in this case.
The null case $z=$constant is simply a null geodesic
of the five dimensional rotating black ring. We are interested in the other solutions which reach
the location of the bulk singularity at the AdS horizon $z=\infty$ for $\rho\to0^-$.

The bulk spacetime has three killing isometries ${\partial}_t,{\partial}_{\psi}$, and ${\partial}_{\phi}$ leading to the corresponding conserved momenta $p_t=-E$, $p_{\psi}=\Psi$ and $p_{\phi}=\Phi$ for geodesic motion. Once again
we consider only those geodesics in the bulk which, on the 4-brane, are restricted to the plane of rotation of the black ring , {\it i.e} in the $x=-1$ plane. The ${g}_{\phi\phi}$ component of the 5+1 dimensional metric goes to zero on the plane of rotation so that $E$ and $\Psi$ are the conserved quantities for such geodesics. The geodesic equations for the $t$ and $\psi$ co-ordinates in the plane of rotation of the black ring are given as
\begin{eqnarray}
\frac{dt}{d\rho}&=&\frac{z^2(1-\lambda)}{l^2(1+{\lambda}y)}\left(\frac{C^2}{(1-\lambda)^2}\frac{(1+y)^4}{(1+{\lambda}y)G(y)}+1\right)E-\frac{z^2C(1+y)^3}{l^2R(1-\lambda)(G(y)}\Psi\nonumber\\ [2mm]
\frac{d\psi}{d\rho}&=& \frac{z^2CR(1+y)^3}{l^2(1-\lambda)G(y)}E-\frac{z^2(1+y)^2(1+{\lambda}y)}{l^2R^2(1-\lambda)G(y)}\Psi\nonumber\\
\end{eqnarray}
The $y$ equation of motion for time like and null geodesics in the bulk which reach the AdS horizon is given by
\begin{equation}
\left( \frac{dy}{d\rho}\right) ^2+\frac{z^4}{l^4}g^{yy}\left( \frac{l^2m^2}{z_1^2}+g^{tt}E^2-g^{t\psi}E{\Psi}+g^{\psi\psi}{\Psi}^2\right) =0.
\end{equation}
Here the contravariant components of the metric in the equation are essentialy the black ring metric without the bulk conformal factor.

The bulk timelike or null geodesics when projected onto the brane reduce to the time like black ring geodesics restricted to the plane of rotation of the ring. The projection to the four brane hypersurface is effected by scaling out the $z$ dependence of the geodesics.  First,
new parameters $\gamma=z^2/m^2\rho$ for null geodesics and $\gamma=(-z_1^2/lm)cot(m\rho/l)$ for time like geodesics are introduced.  We define the rescaled co-ordinates and parameters 
$x=l{\tilde x}/z_1$, $y=l {\tilde y}/z_1$, $t=l {\tilde t}/z_1$
, $R=l^2 {\tilde R}/z_1^2$, $\lambda=z_1\tilde{\lambda}/l$, $\nu=z_1\tilde{\nu}/l$.  The integrals of motion are also rescaled as $E=l {\tilde E}/z_1$,$\Psi=l^3\tilde{\Psi}/z_1^3$.

The geodesic equation for the $y$ coordinate in the rescaled quantities may then be written as,
\begin{equation}
{\left(\frac{d{\tilde{y}}}{d\gamma}\right)}^2 + V_{eff}(\tilde{y};\tilde{E},\tilde{\Psi})=0
\end{equation}
where $V_{eff}$ is the same effective potential as given in eqn. (20).
This is precisely the equation in $y$ for a time like geodesic in the plane of rotation of a five dimensional rotating black ring with an  ADM mass $\tilde {M}$  and angular momentum $\tilde {J}$,
\begin{equation}
\tilde{M}=\left( \frac{z_1}{l}\right)^2 M\,,\qquad\tilde{J}=\left(\frac{z_1}{l} \right)^3J\, 
\end{equation}
and thus existing on the four brane hypersurface located at $z=z_0=l^2/z_1$. The parameter $\gamma$ now serves as the proper time along the time like geodesic.

In order to ascertain the nature of the singularity at the AdS horizon ($z=\infty$) we need to study the behaviour of the bulk geodesics near the AdS horizon, \textit{i.e} as $\rho \rightarrow 0^-$. This is equivalent to $\gamma\rightarrow\infty$, so we need to investigate the late time behaviour of the five dimensional time like geodesics on the four-brane. The geodesics ending into the black ring singularity  will take a finite amount of proper time to do so. For infinite proper time the geodesics can either reach up to the asymptotic infinity on the four brane($\tilde{x},\tilde{y}=-z_1/l$) or remain at a finite distance from the black ring horizon.
The geodesics that reach asymptotic infinity on the brane have late time behaviour
\begin{equation}
\tilde{r} \sim \gamma \sqrt{{\tilde{E}}^2-m^2},
\end{equation} 
where

\begin{equation}
{\tilde{r}}^2=-\frac{1}{z_1/l+\tilde{y}}.
\end{equation}
The co-ordinate $\tilde{r}$ is the radial direction on the brane and it is the same as the radial Myers-Perry coordinate for the black ring in the asymptotic limit modulo certain constants in the plane of rotation of the black ring.

It is expected that stable bound orbits do not exist in the case of the five dimensional black rings. So, only unbound geodesics may reach the AdS horizon at $z\rightarrow \infty$. Along such orbits the curvature squared, Eq.(28), remains finite, thus indicating
the presence of a p-p curvature singularity at the AdS horizon. To explicitly illustrate this, it is necessary to obtain the curvature components in an orthonormal frame parallely propagated on a timelike geodesic to the AdS horizon. Although its simple to demonstrate this in the case of the Schwarzschild black hole in
a braneworld for more complicated metrics and higher dimensions the explicit determination of this frame involves several coupled PDE and renders this analysis computationaly intractable. Although we have to emphasize that such frames exist the choice is
highly non unique and a specific suitable such frame is complicated to establish even for four dimensional Kerr black holes in a braneworld \cite {gs}.
 
\section {Summary and Discussions.} 

To summarize we have described a five dimensional neutral rotating black ring on a four
brane in a six dimensional Randall-Sundrum braneworld.  As mentioned earlier this has been motivated by the fact that for consistency the usual gravitational configurations on the brane, in particular black holes must arise from some higher dimensional bulk solutions. The five dimensional
black ring being the first asymptoticaly flat solution with a non spherical horizon topology is an
interesting configuration to study from a bulk brane world perspective. Especialy as it explicitly 
violates the no hair and the uniqueness theorem. Due to the absence of suitable exact bulk metrics in $D > 4$ a linearized framework around a fixed solution is necessary for the analysis of the black ring in a brane world. In this context the bulk black string approach of Chamblin et. al. \cite {chr} is especialy relevant to elucidate the physical issues although the pathology of a singularity at the AdS horizon persists. However,  absence of such a singularity in lower dimensional brane worlds where exact metrics are available shows the bulk singularity to be a linearization artifact.

To this end we have considered a  bulk six dimensional black string extension of a five dimensional rotating neutral black ring in a 5+1 dimensional Randall-Sundrum braneworld. 
This choice is consistent with the usual reflection symmetric junction conditions on the four brane in such warped compactification models. The bulk black string rotates in the four brane world volume and the induced five dimensional metric on the four brane describes a neutral rotating black ring. This reproduces the usual spacelike curvature singularity of the black ring on the four brane hypersurface. Additionaly a singularity also appears in the bulk at the AdS horizon. After elucidating the geodesics of the rotating black ring restricted to the plane of rotation we have obtained both the timelike and the null geodesics for the black string in the six dimensional bulk. We have further shown that the restricted bulk geodesics projected on the four brane by scaling away the AdS direction exactly match the corresponding class of five dimensional black ring geodesics. The {\it effective potential} has been analysed numericaly and we have shown that stable bound geodesics do n
 ot exist
as is expected in $D > 4$.  It has been further shown that the curvature invariant remains finite along
unbounded geodesics which reach the AdS horizon. This clearly indicates that the bulk curvature singularity at the AdS horizon is possibly a p-p curvature singularity although an explicit illustration using parallely propogated orthonormal frames is computationaly intractable. 

It is mentioned earlier that a fast spinning thin neutral rotating black ring may be described
as a black string boosted along the translationaly invariant direction and identified periodically
in some limit. We have shown that from the bulk perspective this description involves naturaly
a black two brane in the six dimensional bulk orthogonal to the four brane hypersurface. To obtain the black ring on the four brane the black two brane must be boosted along a translationaly invaraint direction longitudinal to the four  brane and identified periodicaly along this direction. Due to the direct equivalence of the two metrics it is obvious that the usual matching of the geodesics on the bulk and the brane will continue to hold in this limit . In the black ring limit the event horzion in the 
bulk would constitute a base $S^2\times S^1$ on the five dimensional brane hypersurface and a trivial $R$ fibration into the bulk.

The issue of stability of the bulk black string configuration is contentious and remains unresolved for axialy symmetric stationary solutions. For AdS solutions one conclusion is that
the prefered phase will be an accumulation of a sequence of lower dimensional black holes  with the
horzion pinched off at some scale. However
for the usual Schwarzschild black string this conclusion has been contested where it has been shown that a more likely scenario is an evolution to a translationaly non invariant stable solution \cite {maeda}. But this although plausible
has not yet been generalized explicitly to axialy symmetric solutions. It has been argued that the bulk solution should pinch off due to the instabilities before reaching the singularity at the AdS horzion \cite {chr, greg1}. However
this issue is far from being completely settled. It is possible that the pathology at the AdS horizon
is a linearization artifact especialy given that lower dimensional exact bulk solutions are regular
everywhere. 

There are several open issues for future studies. Charged rotating black ring solutions have been
obtained in the context of string theory through the $O(d,d)$ transformations. These have been
further generalized to rotating black rings with dipole charges. In the brane world scenario, bulk configurations which reduce to charged black holes have been investigated. It could be shown in this case that the black hole on the brane developed a {\it tidal charge} due to the extra dimensions apart from the usual conserved gauge charge \cite {cbh}. It would be an interesting exercise to study the brane world formulation of the dipole black rings in this context. Very recently it has been shown that in higher diemnsions it is possible to have stable configurations involving combinations of black rings and black holes. These have been christened {\it black saturn} and are remarkably novel solutions of
higher dimensional general relativity \cite {sat}. Naturally it would be interesting to investigate these
configurations from a brane world perspective. It is generaly expected that more such solutions
would be possible in the context of higher dimensions. Some of these issues are being currently studied.

\newpage
\section{Acknowledgements}

We would like to thank A.Virmani for collaboration during
early stages of this work. GS would also like to acknowledge J. Maharana for discussions.
Both of us would like to thank D. D. B. Rao and B. N. Tiwari for computational help.

\end{document}